# Spectrum Sharing for Secrecy Performance Enhancement in D2D-Enabled UAV Networks

Bin Yang, Tarik Taleb, Zhenqiang Wu and Lisheng Ma

*Abstract*—With the assistance of device-to-device (D2D) communications, unmanned aerial vehicle (UAV) networks are anticipated to support widespread applications in the fifth generation (5G) and beyond wireless systems, by providing seamless coverage, flexible deployment, and high channel rate. However, the networks face significant security threats from malicious eavesdroppers due to the inherent broadcast and openness nature of wireless channels. To ensure secure communications of such networks, physical layer security is a promising technique, which utilizes the randomness and noise of wireless channels to enhance secrecy performance. This article investigates physical layer security performance via spectrum sharing in D2D-enabled UAV networks. We first present two typical network architectures where each UAV serves as either a flying base station or an aerial user equipment. Then, we propose a spectrum sharing strategy to fully exploit interference incurred by spectrum reuse for improving secrecy performance. We further conduct two case studies to evaluate the spectrum sharing strategy in these two typical network architectures, and also show secrecy performance gains compared to traditional spectrum sharing strategy. Finally, we discuss some future research directions in D2D-enabled UAV networks.

*Index Terms*—5G-and-beyond, UAVs, D2D, physical layer secrecy performance, spectrum sharing.

## I. INTRODUCTION

Device-to-device (D2D)-enabled unmanned aerial vehicle (UAV) networks (DUAVs) are an important class of emerging network architectures in the fifth generation (5G) and beyond wireless systems, which inherit the distinctive advantages of both UAV and D2D communications [1], [2]. UAVs often have line-of-sight (LoS) channel, providing high-speed connectivity from the sky to different types of ground user equipments (UEs). In addition, they can be deployed on-demand as aerial UEs and flying base stations (BSs) due to their mobility, low cost and high flexibility [3].

However, UAVs also pose many challenges in practical scenarios especially for local congested and disaster areas. In local congested areas, e.g. a stadium, D2D communications can offload heavy traffic from UAVs serving as flying BSs. In large disaster areas, communication range of UAVs cannot cover the whole areas due to their insufficient battery, while D2D communications can extend their wireless coverage by establishing communication links directly. D2D communications have been identified as an efficient supplement of UAV networks to provide proximity based high-speed data delivery services [4], which allow nearby users to share spectrum of cellular networks for accomplishing direct communication bypassing BSs. Therefore, DUAVs are envisioned to support widespread military and civilian applications such as monitoring hostile targets, disaster rescue, firefighting, precision farming, etc. Fig. 1 illustrates a resilient and agile DUAV architecture, aiming to provide various services to satisfy the widespread application requirements. The UAVs can be classified into fixed-wing UAVs and rotary-wing UAVs as shown in Fig. 1. Fixed-wing UAVs have the advantages of longer flight time and faster flight speed, and thus can provide emergency communications in a larger disaster area affected by flooding and earthquake. However, they need a runway to facilitate takeoff and landing. In contrast, the rotary-wing UAVs have shorter flight time and lower speed, but they can take off and land vertically, and hover over particular areas for monitoring fire and for providing local file transfer services in social networks.

The secure communications in DUAVs are challenging due to the broadcast and openness nature of wireless channels. In such networks, transmission information is easily wiretapped by eavesdroppers. Therefore, the information security, especially for financial and military information, is extremely important in the presence of eavesdroppers. The existing security methods mainly adopt upper-layer cryptographic techniques against being intercepted by malicious eavesdroppers, which are often based on the computational complexity. Nevertheless, powerful computing capabilities of devices increase at a very fast speed, such that the cryptographic techniques will be invalid. Moreover, the key management and distribution becomes more challenging in high dynamic DUAVs.

As a compelling remedy, physical layer security (PLS) exploits the inherent characteristics of wireless channels like noise and interference to degrade the eavesdropping channel for guaranteeing secure communications. Recently, PLS has been recognized as a highly attractive approach for providing trustworthy and reliable 5G and beyond wireless communication systems. Cooperative jamming is a typical PLS technique, in which nodes (known as friendly jammers) can inject artificial noise to make the legitimate channel quality advantage over the eavesdropping one. In D2D communications, the important interference issue is caused by spectrum sharing between D2D and cellular UEs, and various methods are

B. Yang is with the School of Computer and Information Engineering, Chuzhou University, Chuzhou, China, and with the School of Electrical Engineering, Aalto University, Espoo, Finland. E-mail: yangbinchi@gmail.com.

T. Taleb is with the School of Electrical Engineering, Aalto University, Finland, with the Information Technology and Electrical Engineering, Oulu University, Finland, and with the Department of Computer and Information Security, Sejong University, South Korea. E-mail: Tarik.Taleb@aalto.fi.

Z. Wu is with the School of Computer Science, Shaanxi Normal University, Xi'an, China. E-mail: zqiangwu@snnu.edu.cn.

L. Ma is with the School of Computer and Information Engineering, Chuzhou University, Chuzhou, China. E-mail: mls@chzu.edu.cn.

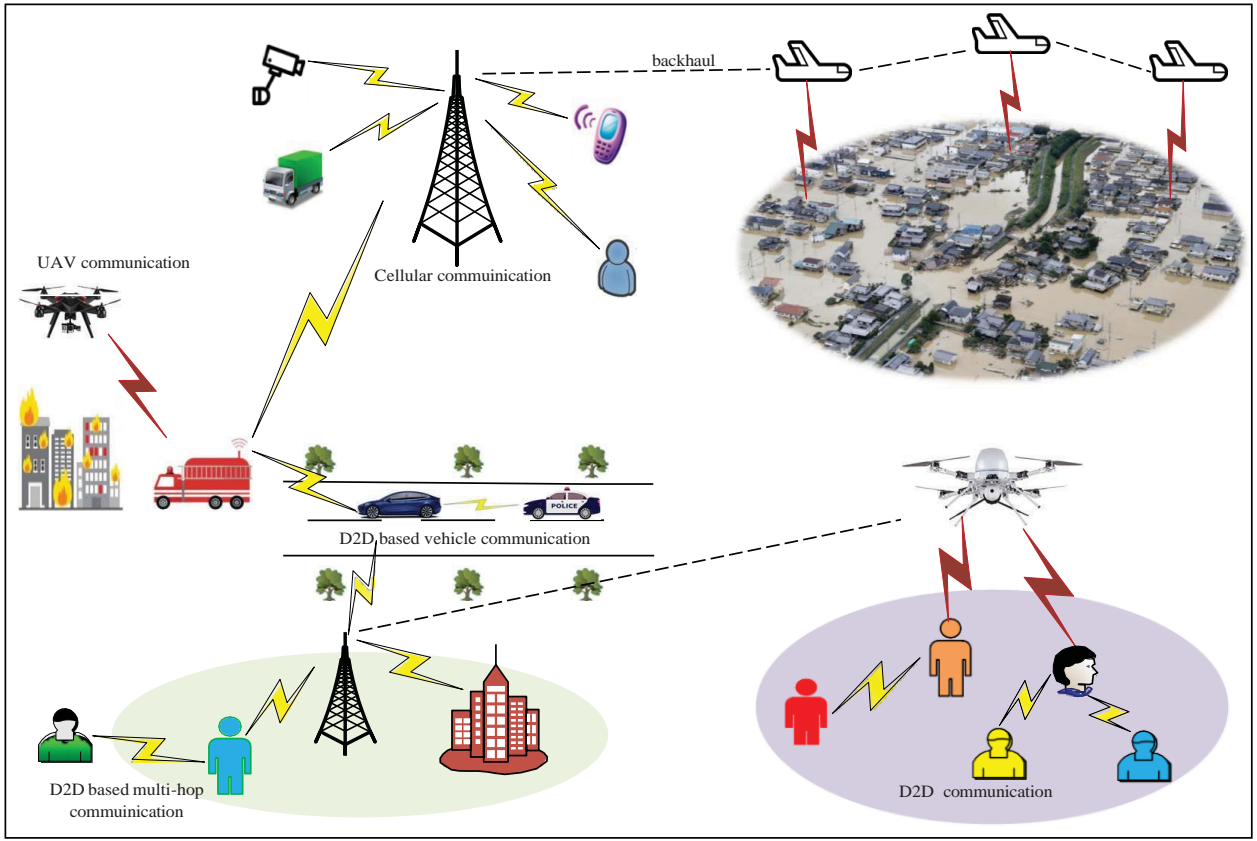

Fig. 1. An example of D2D-enabled UAV networks with high mobility, low cost and on-demand deployment for supporting a range of applications, where each UAV/ground UE can directly communicate with other ground UEs, UAVs and BSs.

utilized to eliminate or mitigate its harmful effect. However, when physical layer security is introduced into D2D communications, the interference can be regarded as similar role of the artificial noise in cooperative jamming to enhance secrecy performance.

Motivated by this consideration, many works were dedicated to the study of secrecy performance in D2D communications from PLS perspective [5]–[9] (see Related Works of Section II-A). These works illustrate the potentials of spectrum sharing for the improvement of secrecy performance. The spectrum sharing in these works can be classified into three categories: cellular, underlay and overlay. Cellular means that cellular UEs use the spectrum of cellular network, underlay is that D2D UEs reuse the spectrum occupied by cellular UEs, and dedicated spectrum is assigned to D2D UEs under overlay. From PLS perspective, the traditional spectrum sharing cannot protect overlay UEs from being intercepted. Moreover, the communications of underlay D2D UEs are insecure once the legitimate channel quality is not better than the eavesdropping one. This is due to the fact that the interference degrades not only the eavesdropping channel but also the legitimate one. To enhance the secrecy performance of DUAVs, one fundamental issue is how to share the spectrum between celluar and D2D UEs. Although many research efforts have been devoted to investigating PLS performance in UAV networks [10]–[14] (see Related Works of Section II-B), these investigations either neglect the importance interference issue or simply consider the spectrum sharing by using the same spectrum for all UEs. In addition, the secrecy performance has not been explored by now in DUAVs. Therefore, a new and dedicated research is deserved to study the spectrum sharing for enhancing PLS performance in DUAVs.

In this article, we introduce PLS to UAV networks with underlaid D2D communications for tackling the challenging information leakage problem in the presence of eavesdroppers. The aim of this article is to achieve secure communications in DUAVs. The spectrum sharing strategy in this article is motived by the traditional one in D2D-enabled cellular networks. The traditional sharing strategy is classified into cellular, underlay and overlay. Specifically, based on this strategy, the authors in [9] consider a selection scheme of spectrum sharing allowing each D2D UE to switch between overlay and underlay patterns with some probability, and a spectrum partition scheme orthogonally partitioning spectrum of cellular networks between cellular and overlay D2D UEs. The corresponding performance of physical layer security (PLS) is further explored under Rayleigh fading channel model. However, the PLS performance of the overlay D2D UEs are not well protected due to the spectrum orthogonally shared among overlay D2D UEs in [9]. Different from the work in [9], we focus on the UAV based network scenario, where UAVs serve as not only flying BSs but also aerial UEs. Meanwhile, we introduce a new underlay cooperative jamming technique, where idle D2D user equipments (UEs) that are not scheduled to send message serve as friendly jammers to generate artificial noise to protect these overlay/underlay UEs. Furthermore, we use a Rayleigh fading channel to model the communication links between ground UEs, and use a Rician fading channel to characterize the line-of-sight (LOS) links from ground UEs to flying BSs. It is notable that our proposed spectrum sharing pattern can well protect the overlay D2D UEs from the attack of the eavesdroppers. In this article, we first present two typical network architectures of UAVs where each UAV acts as either a flying BS or an aerial UE. Then, we introduce a new spectrum sharing strategy to fully utilize interference to enhance secrecy performance of DUAVs, and two case studies are provided to evaluate the new spectrum sharing strategy in these two application scenarios. Finally, we discuss future research directions and conclude this article.

## II. RELATED WORKS

We introduce the related works on the PLS performance study in D2D and UAV communications.

### A. Secure D2D Communications

The existing literature mainly considers the underlay spectrum sharing between cellular and D2D UEs for enhancing PLS performance [5]–[9]. The authors in [5] proposed a theoretical framework using stochastic geometry theory to explore the effect of interference on secrecy performance in D2D underlaid cellular networks. The study finds out that the interference from D2D communications can improve PLS performance of cellular UEs and also bring extra communication opportunities for D2D UEs. A new perspective from the study is provided on the positive effect of the interference in such networks. Following this line, literature [6] proposed a cooperative method with the assistance of coalitional game to examine the secrecy rate under a more general scenario that a D2D UE can reuse multiple spectrum resource blocks and multiple D2D UEs can also reuse the same spectrum resource block. Recently, the authors in [7] investigated the secrecy rate under the following four eavesdropping cases: a single eavesdropper, multiple eavesdroppers, multiple cooperative eavesdroppers and artificial noise, and cooperative colluding eavesdroppers. A joint optimization of power and spectrum allocation of both the D2D and cellular UEs was proposed in [8] to ensure the secrecy communications of cellular UEs and meanwhile improve the spectral utilization of D2D UEs.

### B. Secure UAV Communications

In [10], the authors considered a UAV network consisting of a UAV transmitter, a ground receiver and multiple eavesdroppers, and proposed an optimization algorithm via jointly designing the transmit power and trajectory of the UAV. The algorithm improves the average worst-case secrecy rate of the communication from UAV to ground UE. The authors in [11] further showed the secrecy rate of the communication from UAV to ground UE has more improvement than that from ground UE to UAV by jointly designing the transmit power of UAV and ground UE, and trajectory of UAV. This is because the UAV mobility for the communication from UAV to ground UE affects both the legitimate and eavesdropping channels, rather than only the legitimate channel in the communication from ground UE to UAV. Recently, some research exploited a portion of UAVs as friendly jammers to improve PLS performance in UVA networks [12]–[14]. Specially, in [14], the UAV networks work in millimeter wave frequency band and each UAV acts as not only a UE for information transmission but also a jammer. The literature [14] revealed that the secrecy performance can be significantly improved by optimizing the number of jamming UAVs.

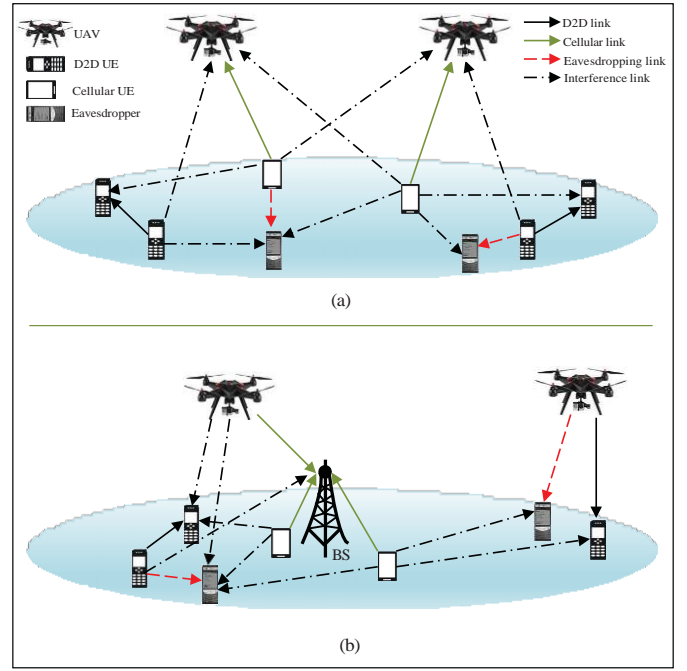

Fig. 2. Illustration of two typical network architectures of DUAVs in the presence of eavesdroppers: a) UAVs serve as flying BSs; b) UAVs serve as aerial UEs that have similar role with ground UEs

## III. NETWORK ARCHITECTURES

In this section, we present two typical network architectures of DUAVs for system models facing security threats, where UAVs serve as either flying BSs or aerial UEs.

### A. Flying BSs for UAVs

The traditional communication infrastructures are (partly) damaged or not available when encountering natural and man-made disasters like earthquake, flooding, war, etc. Accordingly, the emergency data transmissions are extremely critical for disaster relief. To meet such communication requirements, a promising approach is to rapidly deploy UAVs as flying BSs to form cellular network with underlaid D2D communications, which includes the communications from flying BS to flying BS, UE to flying BS, and UE to UE.

In contrast with traditional ground BS assisted wireless communications, the emerging network architecture exhibits many distinctive advantages: (1) it can provide ubiquitous high-speed communication links , especially in disaster areas and remote sensing, since the LoS component dominates the channels from UAVs to ground UEs and proximity based D2D communication channels with high probability, and also the movement of flying BSs in UAV based cellular networks can track the mobility trajectory of ground UEs leading to the decreasing of the intermittence of connectivity; (2) it can also effectively offload traffic from BSs in temporary overloaded hotspots (such as stadiums, concerts and festivals) via direct communication between D2D UEs, and flying BSs and ground UEs; and (3) coverage area of wireless communication can be further extended by deploying multiple flying BSs, and performing D2D communications out of coverage of flying BSs.

Despite the significant gains brought by the network architecture, privacy security is attracting great attention. Specifically, wireless LoS channels are likely to be intentionally listened by some malicious eavesdroppers incurring a risk of privacy information leakage, which seriously threatens the wide deployment of the network for various applications. As shown in Fig. 2a, a cellular network formed by flying BSs is deployed in an area to serve ground UEs in the presence of eavesdroppers. In the network, the cellular UEs can perform uplink transmission with flying BSs, and D2D communication happens between two proximity UEs. The two types of communications could be intercepted by eavesdroppers. The interference generated by spectrum sharing will be exploited to enhance the PLS performance.

### B. Aerial UEs for UAVs

UAVs as aerial UEs are exhibiting great potentials in various Internet of Things (IoT) applications since they have the characteristics of flexible mobility and swift on-demand such that providing ubiquitous connectivity to the IoT. Specially, by carrying onboard some IoT devices, they can be used to target sensing for widespread applications like forest firefighting rescue and human search in the firing high building, where the information of fire/human targets and locations will be sent back to the ground control center to implement real-time target detection and localization. The aerial UEs can also avoid the effect of land traffic roadblocks to provide high-speed air cargo transport services (e.g., Google Wing Project).

It is notable that D2D communications in current network architecture can efficiently enhance performance of UAV networks. For instance, aerial UEs can deliver the same data to numerous ground UEs in a large area, which can be implemented by repeatedly delivering the same data as each aerial UE flies on the different ground UEs. This will lead to the great energy consumption of aerial UEs caused by substantial data retransmissions. To overcome the limitation of their energy, a promising solution is to let ground UEs forward the data with each other by performing direct D2D communications.

However, security threats are hindering the deployment of such a network architecture. As shown in Fig. 2b, a cellular network is deployed in an area to serve ground and aerial UEs, where the malicious eavesdroppers try to listen the channels of cellular and D2D communications. In the network, aerial/ground UEs can perform direct D2D communications with other UEs, and can also communicate with BS as they are close to it. The cellular and D2D communications could be intercepted by eavesdroppers. Spectrum sharing strategy will be applied to the network for enhancing PLS performance.

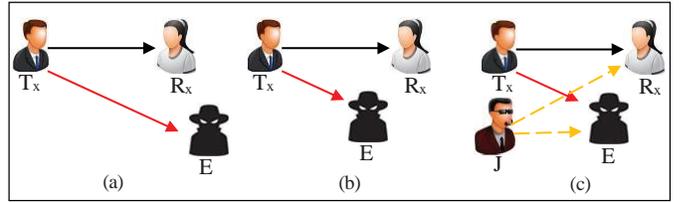

Fig. 3. The transmitter $T_x$ sends data to its desired receiver $R_x$. $E$ and $J$ denote an eavesdropper and a jammer, respectively: a) $R_x$ has a better channel than $E$; b) $E$ has a better channel than $R_x$; c) cooperative jamming. Jammer $J$ generates artificial noise to ensure that $R_x$ has a better channel than $E$.

## IV. SPECTRUM SHARING STRATEGY

We first introduce a typical cooperative jamming based PLS technique which will be used in our spectrum sharing strategy.

### A. Cooperative Jamming

Cooperative jamming represents a class of typical physical layer security techniques that exploit intended or unintended interference as jamming signal to protect the legitimate transmission, as shown in Fig. 3. In Fig. 3a, the quality of main channel between the transmitter $T_x$ and its receiver $R_x$ is better than that of eavesdropping channel between $T_x$ and the eavesdropper $E$. Hence, the secrecy rate of main channel is a positive value, which is defined as the difference of main channel and eavesdropping channel rates, i.e., the maximum achievable rate with perfect secrecy. Note that the secrecy rate is a fundamental metric of PLS performance. In Fig. 3b, the quality of eavesdropping channel is better than that of main channel, leading to zero secrecy rate of main channel. In Fig. 3c, a jammer $J$ generates artificial noise when $T_x$ is sending data to $R_x$. If the quality of the channel between $J$ and $E$ is better than that between $J$ and $R_x$, the jammer $J$ has more effect on the eavesdropper $E$.

### B. Spectrum Sharing Strategy

Traditional spectrum sharing strategy allows each UE to select either cellular, underlay or overlay patterns, whereas none of these patterns may be secure for message transmission in the presence of one eavesdropper. Here, the UEs consist of cellular UEs, and underlay and overlay D2D UEs in the network, where the former ones use the spectrum of cellular pattern, and the latter ones use that of underlay and overlay patterns. In addition, underlay D2D UEs reuse the spectrum of only cellular UEs (as shown in Fig. 4a), such that the message sent by overlay D2D UEs is unprotected from the PLS perspective. Therefore, combining the cooperative jamming technique and underlay pattern, we introduce a new underlay cooperative jamming pattern, where idle D2D UEs that do not send message serve as friendly jammers to generate artificial noise to protect these UEs reusing the same spectrum with them. We further propose a new spectrum sharing strategy as shown in Fig. 4b, which allows D2D UEs in underlay/underlay cooperative jamming patterns to reuse the spectrum of cellular UEs/overlay D2D UEs so as to provide a full PLS-based security solution.

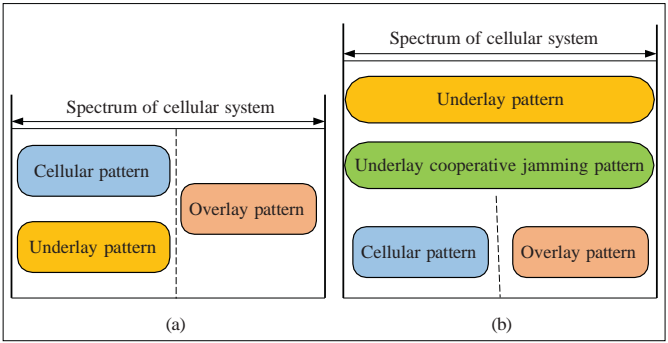

Fig. 4. Spectrum sharing strategy: a) traditional spectrum sharing strategy; b) new spectrum sharing strategy.

As shown in Fig. 4b, the total spectrum of cellular system is divided into two portions. We use $\eta$ to denote spectrum partition factor. A fraction $\eta$ of total spectrum is orthogonally and equally assigned to each cellular UE and the remaining fraction $(1-\eta)$ is assigned to each overlay D2D UE in the same way. A signal can be successfully decoded if and only if the SINR of the received signal at a BS or UE is greater than some threshold $\beta$. Otherwise, the transmitting UE keeps idle. Each idle D2D serves as a potential friendly jammer to generate artificial noise for enhancing secrecy rate performance. For instance, consider a simple network scenario with a transmitter $T_x$, a receiver $R_x$, an eavesdropper $E$ and a jammer $J$ as shown in Fig. 3c. When jamming signal is injected into such a network, the secrecy rate $SR_{T_xR_x}$ can be determined as

$$SR_{T_xR_x} = B \left[ \log_2 \frac{1+\text{SINR}_{T_xR_x}}{1+\text{SINR}_{JR_x}} - \log_2 \frac{\text{SINR}_{T_xE}}{1+\text{SINR}_{JE}} \right]^+ \quad (1)$$

where $B$ denotes the bandwidth of the channel from $T_x$ to $R_x$, $[x]^+ = \max(0, x)$, $\text{SINR}_{T_xR_x}$, $\text{SINR}_{T_xE}$, $\text{SINR}_{JR_x}$ and $\text{SINR}_{JE}$ denote the signal-to-interference-plus-noise ratio (SINR) at $R_x$, $E$, $R_x$ related to $J$, $E$ related to $J$, respectively. We can see that the secrecy rate $SR_{T_xR_x}$ can be improved only if the jammer $J$ hurts the eavesdropper $E$ more than the receiver $R_x$. To ensure the system performance, we select these strong idle D2D UEs as friendly jammers to generate artificial noise, i.e., for each strong idle D2D UE $J$, the $\text{SINR}_{JR_x}$ at the receiver of $J$ is less than the $\text{SINR}_{JE}$ at the eavesdropper $E$.

## V. CASE STUDIES

The goal of study is to evaluate the new spectrum sharing strategy for secrecy rate performance in these two general network architectures as shown in Fig. 2. On one hand, a swarm of UAVs as flying BSs to provide wireless service to a large number of ground UEs, and these UEs can also perform D2D communications for reducing traffic load of flying BSs and extending coverage area as shown in Fig. 2a, where a group of eavesdroppers are trying to intercept information from cellular and D2D communications. On the other hand, ground UEs and UAVs as aerial UEs perform either cellular or D2D communications. Under these two cases, idle D2D UEs can serve as friendly jammers to generate artificial noise.

In this study, each UE selects its communication mode according to received signal strength (RSS)-based mode selection: cellular communication mode is used if the RSS of its closest (flying) BS is more than that of its closest (aerial) UE; otherwise, D2D communication mode is selected. The ground BSs, UAVs, UEs and eavesdroppers are randomly distributed in a three dimensional space according to homogeneous Poisson point processes with densities $\lambda_B$, $\lambda_A$, $\lambda_U$, and $\lambda_E$. We use PPP to model the locations' distribution of BSs, UEs, UAVs and eavesdroppers. This is because PPP can well characterize the random locations of mobile users. Specially, it has been proven that PPP can provide high accurate for modeling the practical deployment of BSs. Each (aerial) UE works on one of cellular and D2D communication modes according to the RSS-based mode selection. The D2D UEs are either underlay D2D UEs or overlay D2D UEs with equal probability. Each underlay D2D UE selects to reuse the spectrum of one cellular UE with probability $\eta$ and that of one overlay D2D UE with probability $1-\eta$. Here, we only consider the rotary-wing UAVs hovering over the targeted area with altitude $H$.

We focus on a given D2D pair and a cellular UE. The basic idea of the simulation for the secrecy rate performance is included in the following Algorithm 1 regarding our proposed spectrum sharing strategy and the scenario of UAVs as flying BSs. Similarly, we can also obtain the secrecy rate performance under the scenario of UAVs as aerial BSs.

TABLE I
SIMULATION PARAMETERS

| Parameter | Value under UAVs as flying BSs | Value under UAVs aerial UEs |
|---|---|---|
| Network area $S$ | $10^6$ m$^2$ | $10^6$ m$^2$ |
| Total system bandwidth $W$ | 2 GHz | 2 GHz |
| Density of BSs $\lambda_B$ | 0 | $4 \times 10^{-5}$ BSs/m$^2$ |
| Density of UAVs $\lambda_U$ | $10^{-4}$ UAVs/m$^2$ | Varying from $10^{-3}$ to $5.5 \times 10^{-3}$ UAVs/m$^2$ |
| Density of UEs $\lambda_A$ | 0.2 UEs/m$^2$ | 0.01 UEs/m$^2$ |
| Density of eavesdroppers $\lambda_E$ | Varying from 0.001 to 0.154 eavesdroppers/m$^2$ | 0.098 eavesdroppers/m$^2$ |
| Flying altitude of UAVs $H$ | 300 m | 200 m |
| Transmit power of UAVs $P_A$ | 200 mW | 200 mW |
| Transmit power of UEs $P_U$ | 230 mW | 300 mW |
| Received signal threshold $\beta$ | −120 dBm | −120 dBm |
| Spectrum partition factor $\eta$ | 0.6 | 0.5 |
| Noise variance $\sigma^2$ | -130 dBm | -130 dBm |
| Path loss exponents $\alpha_a$ | 2 for the channel between ground UEs and flying BSs | 2 for the channel between aerial UEs /ground UEs and aerial UEs |
| Path loss exponents $\alpha_g$ | 4 for the channel between ground UEs | 4 for the channel between ground UEs |

---

**Algorithm 1** Secrecy rate performance under the scenario of UAVs as flying BSs:

1. **Input:** These parameters defined in Table 1.
2. **Output:** The secrecy rate of overlay D2D link; the secrecy rate of cellular link.
3. **Initialize:** The total number of UAVs: $S\lambda_A$, the total number of UEs: $S\lambda_U$.

    1) We first differentiate D2D UEs and cellular UEs according to the following mode selection: For each transmitting UE, if the received signal strength (RSS) at its closest flying BS is more than that at its closest UE, then it is a cellular UE; otherwise, it is a D2D UE.

    2) Since each D2D UE is either overlay one or underlay one with equal probability, we further differentiate overlay ones and underlay ones. Meanwhile, we can also determine the number of cellular UEs, underlay D2D UEs and overlay D2D UEs.

    3) We find idle D2D UEs among all D2D ones according to the following rule: If the SINR at the receiver of a D2D UE is less than $\beta$, then the D2D UE keeps idle. Each idle D2D UE is a potential friendly jammer.

    4) If the SINR of jamming signal from a potential friendly jammer at the receiver of the given overly D2D UE (or the given cellular UE) is less than the SINR of the jamming signal at the eavesdropper of the receiver, then the potential friendly jammer can transmit noise to protect the given overlay D2D UE (or the given cellular UE) from being intercepted by the eavesdropper.

    5) We determine the SINR at the receiver of the given overlay D2D UE, and the SINR at the receiver of the given cellular UE. The bandwidth of link from the overly D2D UE to its receiver is determined as $W(1-\eta)$ divided by the number of overlay D2D UEs, and the bandwidth of link from the given cellular UE to its destined flying BS is determined as $W\eta$ divided by the number of cellular UEs.

    6) Based on the results from 5), we can obtain the secrecy rate of overlay D2D UE and that of cellular UE.

---

### A. Numerical Results Under the Case of Flying BSs

In this section, we conduct simulation study to evaluate the new spectrum sharing strategy according to the performance metric of secrecy rate in the scenario with UAVs as flying BSs shown in Fig. 2a. We further compare the secrecy performance of the new spectrum sharing strategy with that of the traditional one in Fig. 4. We use a Rayleigh fading channel to model both small scale and large scale fading for the communication links between ground UEs, and use a Rician fading channel to characterize the LOS links from ground UEs to flying BSs [15].

We summarize in Fig. 5 the impacts of the density of eavesdropper $\lambda_E$ on the secrecy rates of overlay D2D link and cellular link under the new and traditional spectrum sharing strategies. We can see from Fig. 5 that both the secrecy rates of overlay D2D and cellular links decrease as $\lambda_E$ increases. This phenomenon can be explained as follows. In our study, we consider that each eavesdropper wiretaps the message from the transmitter closest to it. Thus, the distance between the eavesdropper and corresponding transmitter decreases as $\lambda_E$ increases, which leads to the increasing of the rate of the eavesdropping link. According to the definition of secrecy rate, the secrecy rate decreases with the increasing of $\lambda_E$. We further observe that under the traditional spectrum sharing strategy, as $\lambda_E$ increases beyond some constant, the secrecy rate becomes zero, which implies that the transmission message over both the overlay D2D and cellular links can be wiretapped by eavesdroppers. It is notable that the values of secrecy rate are positive under the new spectrum sharing strategy, due to the fact that underlay D2D UEs reusing cellular/overlay spectrum and friendly jammers can protect the cellular and overlay D2D transmissions against wiretapping. It indicates that the new spectrum sharing strategy can enhance secrecy performance of DUAVs in comparison with the traditional one.

### B. Numerical Results Under the Case of Aerial UEs

We continue to conduct simulation study in the scenario with UVAs as aerial UEs. We summarize in Fig. 6 that the impacts of the number of UAVs on the secrecy rates of overlay D2D and cellular links under the new and traditional spectrum sharing strategies. It can be observed from Fig. 6 that both the secrecy rates of cellular and overlay D2D links increase with the number of UAVs. Recall that each UAV can select cellular or D2D communication modes according to RSS-based mode selection. As the number of UAVs increases, the distances of cellular and overlay D2D links decrease, which leads to the increasing of secrecy rates. Another careful observation from Fig. 6 also indicates that the secrecy performance under the new spectrum sharing strategy is better than that under the traditional one. Specially, under the traditional one, malicious eavesdroppers can intercept the transmission message under the case of the zero secrecy rate as shown in Fig. 6.

## VI. FUTURE RESEARCH DIRECTIONS

**Malicious spoofing attack:** DUAVs may face eavesdropping as well as malicious spoofing attack, due to broadcast and openness nature of wireless medium. Malicious attackers may launch more advance spoofing attacks to send forged message to receiver, which brings huge security threats in some critical areas, like D2D-based vehicle communication. Therefore, it is desired to conduct a more in-depth study of the spoofing attack issue in DUAVs.

**Covert UAV and D2D communications:** PLS technique is in the sense that adversaries cannot correctly decode the signal received by them, but it cannot prevent adversaries from detecting transmission signal from legitimate transmitters. Specially, in military area, once if adversaries successfully detect the signal transmitted by UAVs or ground devices, these UAVs and ground devices are probably attacked. To overcome this problem, a promising solution is covert communication technique, which aims to hide the very existence of wireless transmissions from the adversaries. Thus, it can provide a strong security guarantee for transmitters to prevent the attacks of adversaries for supporting many applications, like covert military communications, location tracking and intercommunication of Internet of Things [16]–[19]. For the purpose of further understanding the meaning of covert communication, we consider a simple covert communication system consisting of a transmitter, a receiver and an adversary. The transmitter is transmitting sensitive information to the receiver, while the adversary tries to detect the transmission from the transmitter to its receiver. The transmitter can conceal the information in artificial noise to prevent the adversary from knowing the information transmission process.

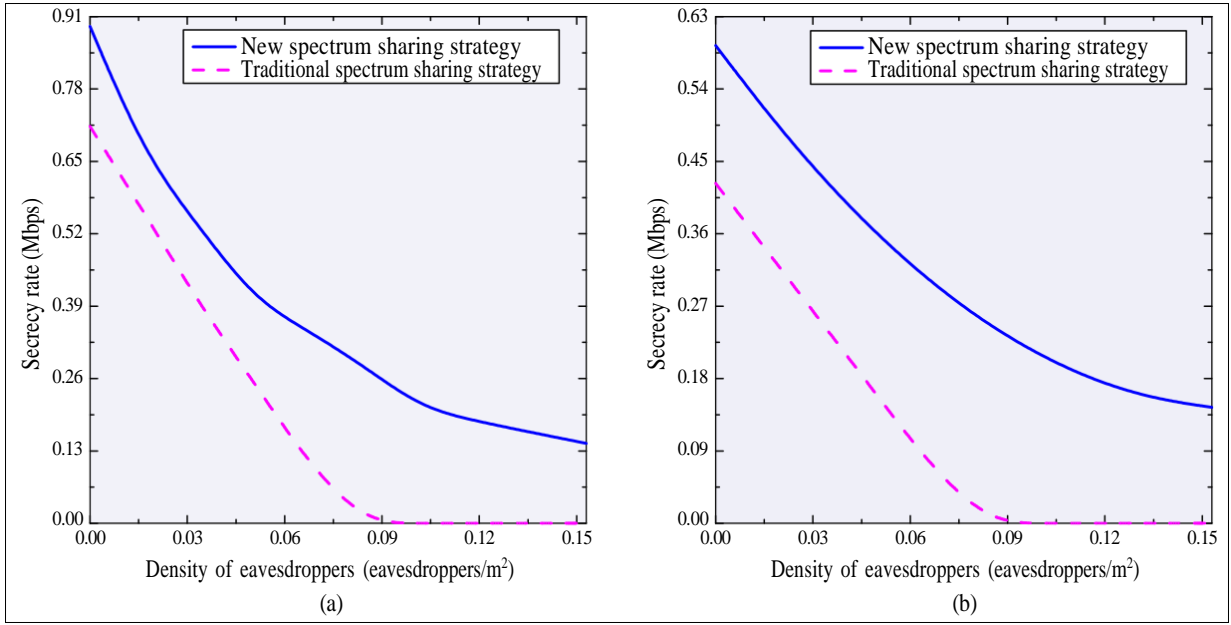

Fig. 5. Secrecy rate in the scenario with UAVs as flying BSs: a) secrecy rate of overlay D2D link; b) secrecy rate of cellular link.

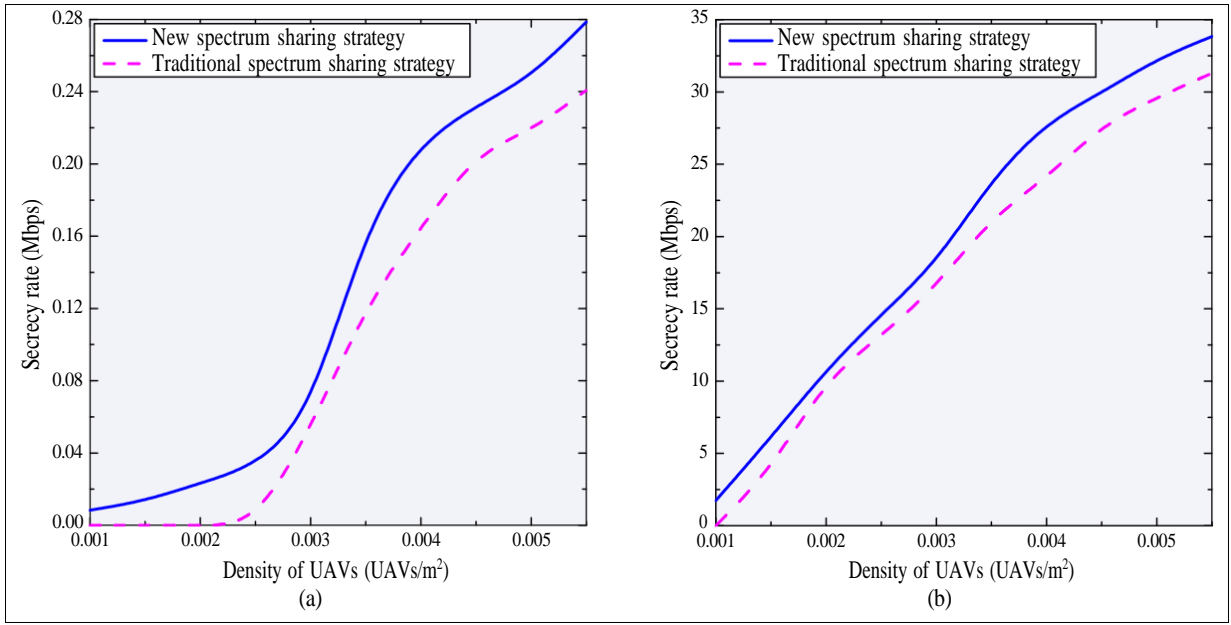

Fig. 6. Secrecy rate in the scenario with UAVs as aerial UEs: a) secrecy rate of overlay D2D link; b) secrecy rate of cellular link.

**Millimeter wave based UAV and D2D communications:** Millimeter wave is identified as one of the most promising technologies to meet the demanding data traffic requirements in the 5G and beyond, since it can provide a large amount of spectrum available from 30 to 300GHZ for supporting high-speed data transmission. Millimeter wave with directional antennas and large bandwidths can offer rich opportunities for DUAVs. To guarantee secure communication in millimeter wave based DUAVs, it is deserved to study the security issues from physical layer security and covert communication perspective.

## VII. CONCLUSIONS

Spectrum sharing is an important solution to increase PLS secrecy performance of DUAVs. Different from traditional spectrum sharing strategy, this article presents a new spectrum sharing strategy that can fully utilize interference of communication links to guarantee the security of cellular and D2D communications in DUAVs. We further evaluate our spectrum sharing strategy under two typical application scenarios, where UAVs serve as flying BSs and aerial UEs, respectively. The corresponding numerical results illustrate that our spectrum sharing strategy can significantly enhance the secrecy rate performance of DUAVs in comparison with traditional one, and provide a full PLS-based security solution for supporting wide-ranging applications. Finally, we shed light on some future research directions. Besides, an interesting future work is to conduct the theoretical analysis on the new spectrum sharing strategy using stochastic geometry.

## ACKNOWLEDGEMENT

This work was partially supported by the European Union's Horizon 2020 Research and Innovation Program through the 5G!Drones Project under Grant No. 857031, the Academy of Finland 6Genesis project under Grant No. 318927, the Academy of Finland CSN project under Grant No. 311654, the NSF of China under Grant No. 61702068 and 61962033, the Anhui Province project under Grant No. 1808085MF165, gxgwfx2019060 and KJ2019A0643, the Yunnan Province project under Grant No. 2018FH001-010, and the Chuzhou University project under Grant No. zrjz2019011.

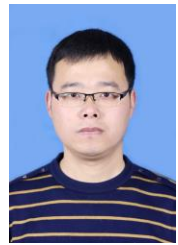

**Bin Yang** received the B.S. and M.S. degrees both in computer science from Shihezi University, China, in 2004 and from China University of Petroleum, Beijing Campus, in 2007, and Ph.D. degree in systems information science from Future University Hakodate, Japan in 2015, respectively. He is currently an associate professor with the School of Computer and Information Engineering, Chuzhou University, China, and is also a senior researcher with the School of Electrical Engineering, Aalto University, Finland. His research interests include unmanned aerial vehicles, D2D communications, cyber security and Internet of Things.

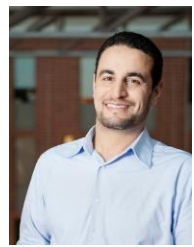

**Tarik Taleb** received the B.E. degree (with distinction) in information engineering in 2001, and the M.Sc. and Ph.D. degrees in information sciences from Tohoku University, Sendai, Japan, in 2003, and 2005, respectively. He is currently a Professor with the School of Electrical Engineering, Aalto University, Espoo, Finland. He is the founder and the Director of the MOSA!C Lab. He is the Guest Editor-in-Chief for the IEEE JSAC series on network softwarization and enablers.

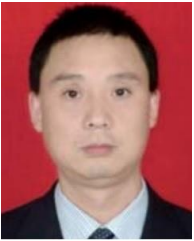

**Zhenqiang Wu** received the B.S. degree from Shaanxi Normal University, China, in 1991, and the M.S. and Ph.D. degrees from Xidian University, China, in 2002 and 2007, respectively. He is currently a Full Professor with Shaanxi Normal University, China. His research interests include computer communications networks, mainly wireless networks, cyber security, anonymous communication and privacy protection.

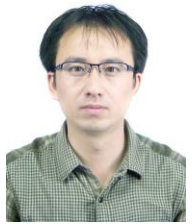

**Lisheng Ma** received the B.S., M.S. and Ph.D. degrees from Taiyuan Normal University, China in 2004, from Southwest University, China in 2007 and from Future University Hakodate, Japan in 2017, respectively. He is currently an associate professor of Chuzhou University, China. His research interests include cyber security, switching networks and data center networks.